\documentclass[fleqn,twoside]{article}
\usepackage{espcrc2}

\usepackage{graphicx}
\usepackage{amsmath}
\usepackage{amssymb}

\usepackage[figuresright]{rotating}

\title{Physics Potential of Future Atmospheric Neutrino Searches}

\author{Thomas Schwetz\address{Max-Planck-Institute for Nuclear Physics,
PO Box 103980, 69029 Heidelberg, Germany}}
\begin{document}     

\begin{abstract}
The potential of future high statistics atmospheric neutrino
experiments is considered, having in mind currently discussed huge
detectors of various technologies (water Cerekov, magnetized iron,
liquid Argon). I focus on the possibility to use atmospheric data to
determine the octant of $\theta_{23}$ and the neutrino mass
hierarchy. The sensitivity to the $\theta_{23}$-octant of atmospheric
neutrinos is competitive (or even superior) to long-baseline
experiments. I discuss the ideal properties of a fictitious
atmospheric neutrino detector to determine the neutrino mass
hierarchy.
\vspace{1pc}
\end{abstract}

\maketitle

\section{Introduction}

In the past atmospheric neutrinos have played an important role in
establishing the phenomenon of neutrino oscillations via observing the
dominant oscillation mode in the 2-3 sector of Lepton mixing
\cite{SKatm}.  However, very soon precision long-baseline (LBL)
experiments \cite{lbl,t2k,Huber:2004ug} will outperform
atmospheric neutrinos in the determination of the oscillation
parameters $|\Delta m^2_{31}|$ and $\sin^22\theta_{23}$. Hence, the
question {\it ``Can we learn something more from atmospheric neutrinos
in the era of precision neutrino oscillation experiments?''}
arises. In the following I will suggest that the answer to this
question is ``yes'', by considering sub-leading three-flavour effects
(see, e.g., \cite{Akhmedov:2008qt} and references therein)
in future high statistics atmospheric neutrino experiments.  Projects
currently under discussion \cite{laguna} include Mt scale water
\v{C}erenkov detectors~\cite{Wcerenkov}, large magnetized iron
calorimeters~\cite{INO}, or 100 kt scale liquid Argon time projection
chambers \cite{lar}. See also~\cite{Choubey:2006jk} for a recent
review.

\section{Combining LBL and ATM data from Mt water detectors}

The primary aims of future neutrino experiments are the determination
of the mixing angle $\theta_{13}$, the CP-phase $\delta_\mathrm{CP}$,
and the type of the neutrino mass hierarchy (normal or inverted),
i.e., the sign of $\Delta m^2_{31}$. It is well known that parameter
degeneracies are a severe problem on the way towards these goals. In
Ref.~\cite{Huber:2005ep} it was demonstrated that for LBL experiments
based on Mt scale water \v{C}erenkov detectors data from atmospheric
neutrinos (ATM) provide an attractive method to resolve degeneracies.

Atmospheric neutrinos are sensitive to the neutrino mass hierarchy if
$\theta_{13}$ is sufficiently large due to Earth matter effects,
mainly in multi-GeV $e$-like events~\cite{atm13}, see, e.g.,
\cite{Gandhi:2007td} for recent studies. Moreover, sub-GeV $e$-like
events provide sensitivity to the octant of
$\theta_{23}$~\cite{Peres:2003wd,Gonzalez-Garcia:2004cu} due to
oscillations with $\Delta m^2_{21}$. However, these effects can be
explored efficiently only if LBL data provide a very precise
determination of $|\Delta m^2_{31}|$ and $\sin^22\theta_{23}$, as well
as some information on $\theta_{13}$~\cite{Huber:2005ep}.

\begin{figure*}[!ht]
\centering
\includegraphics[width=0.81\textwidth]{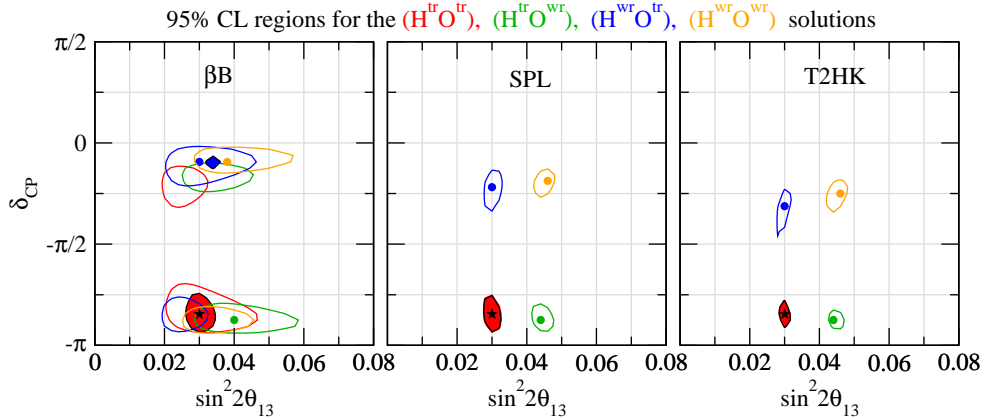}
\vspace*{-6mm}
\caption{Allowed regions in $\sin^22\theta_{13}$ and
  $\delta_\mathrm{CP}$ for LBL data alone (contour lines) and LBL+ATM
  data combined (colored regions). The true parameter values are
  $\delta_\mathrm{CP} = -0.85 \pi$, $\sin^22\theta_{13} = 0.03$,
  $\sin^2\theta_{23} = 0.6$. $\mathrm{H^{tr/wr} (O^{tr/wr})}$
  refers to solutions with the true/wrong mass hierarchy (octant of
  $\theta_{23}$).}
\label{fig:degeneracies}
\end{figure*}
 
Here we illustrate the synergies from a combined LBL+ATM analysis at
the examples of the T2K phase II experiment~\cite{t2k} (T2HK) with the
HK detector of 450~kt fiducial mass, and two experiments with beams
from CERN to a 450~kt detector at Frejus (MEMPHYS)\cite{c2m}, namely
the SPL super beam and a $\gamma = 100$ beta beam ($\beta$B). The LBL
experiments are simulated with the GLoBES software~\cite{globes}, and a
general three-flavor analysis of ATM data is
performed~\cite{c2m,Huber:2005ep,Gonzalez-Garcia:2004cu}. For each
experiment we assume a running time of 10 years, where the
neutrino/anti-neutrino time is chosen as 2+8 years for SPL and T2K, and
5+5 years for the beta beam, see~\cite{c2m} for details.

The effect of degeneracies becomes apparent in
Fig.~\ref{fig:degeneracies}. For given true parameter values the data
can be fitted with the wrong hierarchy and/or with the wrong octant of
$\theta_{23}$. Hence, from LBL data alone the hierarchy and the octant
cannot be determined. Moreover, as visible from the solid lines in
Fig.~\ref{fig:degeneracies} the degenerate solutions appear at
parameter values different from the true ones, an hence, ambiguities
exist in the determination of $\theta_{13}$ and $\delta_\mathrm{CP}$.
If the LBL data are combined with ATM data only the colored regions in
Fig.~\ref{fig:degeneracies} survive, i.e., in this particular example
for all three experiments the degeneracies are completely lifted at
95\%~CL, the mass hierarchy and the octant of $\theta_{23}$ can be
identified, and the ambiguities in $\theta_{13}$ and
$\delta_\mathrm{CP}$ are resolved. Let us note that here we have
chosen a favorable value of $\sin^2\theta_{23} = 0.6$; for values
$\sin^2\theta_{23} < 0.5$ in general the sensitivity of ATM data is
weaker~\cite{Huber:2005ep}.

\begin{figure}[!tb]
\centering
\includegraphics[width=0.43\textwidth]{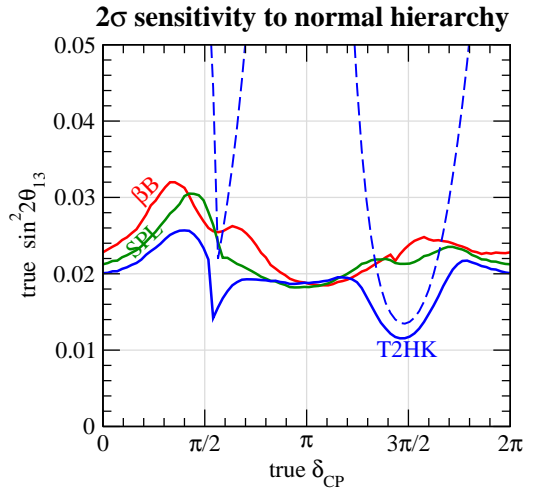}
\vspace*{-6mm}
\caption{Sensitivity to the neutrino mass hierarchy as a function of
   $\sin^22\theta_{13}$ and $\delta_\mathrm{CP}$ for
   $\theta_{23}^\mathrm{true} = \pi/4$ and a true normal
   hierarchy. Solid curves correspond to LBL+ATM data combined, the
   dashed curves correspond to T2HK LBL data-only. $\beta$B and SPL
   without ATM have no sensitivity to the hierarchy.}
\label{fig:hierarchy}
\end{figure}

In Fig.~\ref{fig:hierarchy} we show the sensitivity to the neutrino
mass hierarchy. For LBL data alone there is practically no sensitivity
for the CERN--MEMPHYS experiments (because of the very small matter
effects due to the relatively short baseline of 130~km), and the
sensitivity of T2HK depends strongly on the true value of
$\delta_\mathrm{CP}$. However, with the LBL+ATM combination all
experiments can identify the mass hierarchy at $2\sigma$~CL provided
$\sin^22\theta_{13} \gtrsim 0.02-0.03$~\cite{c2m}.

\begin{figure}[!tb]
\centering
  \includegraphics[width=0.4\textwidth]{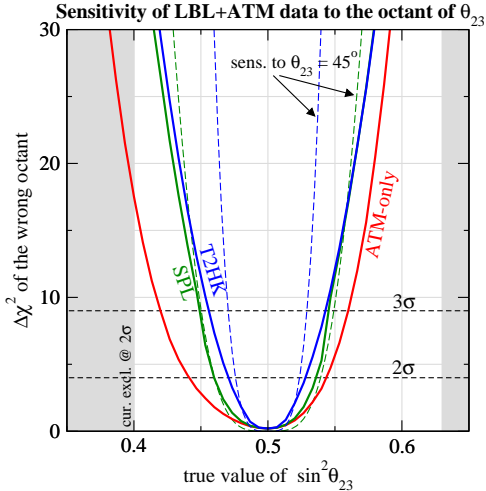}
\vspace*{-6mm}
  \caption{$\Delta\chi^2$ of the solution with the wrong octant of
  $\theta_{23}$ as a function of $\sin^2\theta_{23}$. We have assumed
  a true value of $\theta_{13} = 0$. The dashed curves show the
  $\Delta\chi^2$ of $\theta_{32} = 45^\circ$, i.e., the ability to
  exclude maximal mixing.}
  \label{fig:octant}
\end{figure}

Fig.~\ref{fig:octant} shows the potential of ATM+LBL data to exclude
the octant degenerate solution. Since this effect is based mainly on
oscillations with $\Delta m^2_{21}$ there is very good sensitivity
even for $\theta_{13} = 0$; a finite value of $\theta_{13}$ in general
improves the sensitivity\cite{Huber:2005ep}.  From the figure one can
read off that atmospheric data alone can resolve the correct
octant at $3\sigma$ if $|\sin^2\theta_{23} - 0.5| \gtrsim 0.085$. If
atmospheric data is combined with the LBL data from SPL or T2HK there
is sensitivity to the octant for $|\sin^2\theta_{23} - 0.5| \gtrsim
0.05$. 

\section{Magnetized iron calorimeters}

In water \v{C}erenkov detectors one cannot distinguish between
neutrino and anti-neutrino events. This limits the sensitivity to the
mass hierarchy, since depending on the hierarchy the resonance occurs
either for neutrinos or anti-neutrinos. Therefore, in principle one
expects that the sensitivity improves for detectors capable to
distinguish atmospheric neutrino from anti-neutrino events. In the
following we discuss the possibility offered by a large (several
10~kt) magnetized iron calorimeter similar to the INO
proposal\cite{INO}. Such a detector can determine the charge of muons,
whereas electron detection is difficult. The principles of atmospheric
neutrino measurements with a 5.4~kt detector of this type have been
established recently by the MINOS experiment\cite{Adamson:2005qc}.

Here we report the results obtained in Ref.\cite{Petcov:2005rv}, see
Ref.~\cite{magn} for related considerations.  We limit ourselves to
$\mu$-like events, and we assume a correct identification of
$\nu_\mu$- versus $\bar\nu_\mu$-events of 95\%. The observation of the
muon and the hadronic event allows in principle to reconstruct the
original direction and energy of the neutrino. Indeed, it has been
stressed in Ref.~\cite{Petcov:2005rv} that the accuracy of neutrino
energy and direction reconstruction is crucial for the determination
of the hierarchy. The reason is that the difference in the event
spectra of normal and inverted hierarchy show a characteristic
oscillatory pattern. If this pattern can be resolved a powerful
discrimination between the hierarchies is possible. If however, the
oscillatory pattern is washed out because of a poor accuracy in energy
and direction reconstruction the sensitivity to the hierarchy
decreases drastically.

\begin{figure*}[bt]
\centering
  \includegraphics[width=0.75\textwidth]{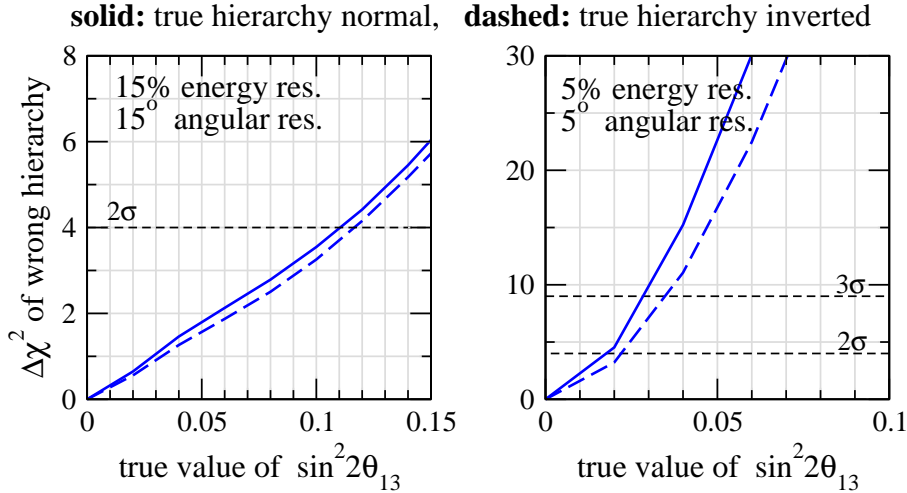}
\vspace*{-6mm}
  \caption{Sensitivity to the neutrino mass hierarchy of a magnetized
  iron calorimeter for different assumptions on the neutrino energy
  and direction reconstruction accuracy. We assume 500 kt yr data from
  an INO-like detector, corresponding to 4000 up-going $\mu$-like
  events with $E_\mu > 2$~GeV. Here, $\theta_{23}^\mathrm{true} =
  \pi/4$ and we assume external information on $|\Delta m^2_{31}|$ and
  $\sin^22\theta_{23}$ of 10\%, and an uncertainty on
  $\sin^22\theta_{13}$ of $\pm 0.02$.}
  \label{fig:magn}
\end{figure*}

In Fig.~\ref{fig:magn} we show the sensitivity to the hierarchy for a
500~kt~yr exposure of an INO-like detector. In the left panel we
assume that the neutrino energy can be reconstructed with an accuracy
of 15\% and the neutrino direction with an accuracy of $15^\circ$,
whereas in the right panel the very optimistic accuracies of 5\% and
$5^\circ$ are adopted. Details on our simulation and systematic errors
are given in Ref.~\cite{Petcov:2005rv}. One observes from the plot that
for optimistic assumptions the hierarchy can be identified at
$2\sigma$ if $\sin^22\theta_{13} \gtrsim 0.02$. This sensitivity is
comparable to the one from Mt water \v{C}erenkov detectors discussed in
the previous section. If however, more realistic values for the energy
and direction reconstruction are adopted the sensitivity deteriorates
drastically and values of $\sin^22\theta_{13} \gtrsim 0.1$ (close to
the present bound) are required.

\section{The ideal atmospheric neutrino detector}

\begin{figure*}[t]
\centering
  \includegraphics[width=0.65\textwidth]{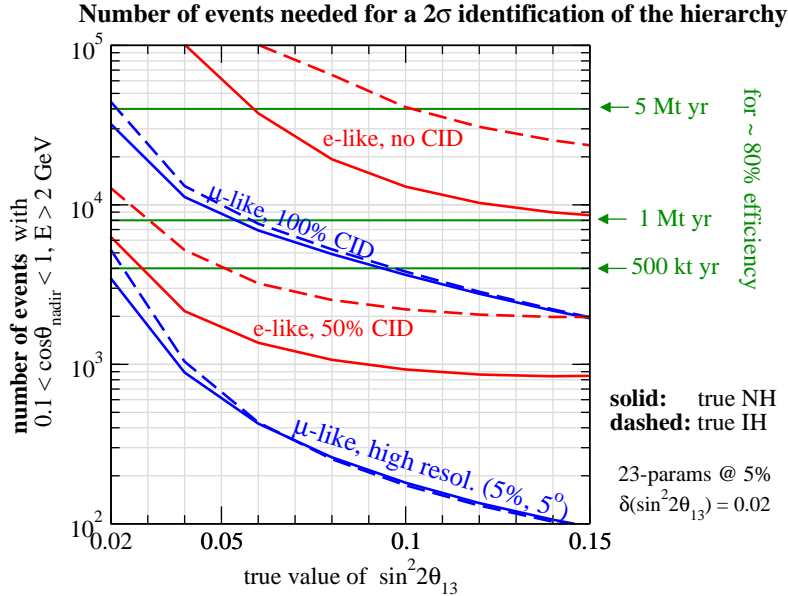}
\vspace*{-6mm}
  \caption{
  Number of up-going events ($\cos\theta_\mathrm{nadir} >
  0.1$), with $E_\nu > 2$~GeV for a $2\sigma$ determination of the
  neutrino mass hierarchy. Curves are shown for different assumptions
  on event types ($e$- vs $\mu$-like), charge identification (CID),
  and energy and direction reconstruction. The horizontal lines
  indicate the typical exposures (fiducial mass times time) assuming
  80\% efficiency.}
  \label{fig:which}
\end{figure*}

What are the properties of an ideal atmospheric neutrino detector to
determine the mass hierarchy? In Fig.~\ref{fig:which} I show the
number of events needed to obtain a hierarchy determination at
$2\sigma$ for various assumptions on the event sample. One can
summarize the results in the following way. An ideal detector 
should be able to
\begin{itemize}
\item
see $e$-like events with charge ID (at least statistically),
\item\vspace*{-2mm}
see $\mu$-like events with charge ID,
\item\vspace*{-2mm}
reconstruct neutrino energy (direction)
at the level of few \% (degree)
for $\mu$ like events,
\item\vspace*{-2mm}
and it should be very big.
\end{itemize}

Let us emphasize that for an $e$-like event sample even a rather
modest statistical separation of neutrino from anti-neutrino events
will improve drastically the sensitivity. The example with 50\% CID
shown in Fig.~\ref{fig:which} assumes that $e$-like events can be
separated into two samples with $\nu/\bar\nu$ ratios of 2:1 and 1:2,
respectively. Possibilities of statistical separation of neutrino and
anti-neutrino events without a magnetic field have been discussed
recently in the context of a low-energy Neutrino Factory
\cite{Huber:2008yx}. The possibility to explore the different
fractions of single and multi-ring events for neutrinos and
anti-neutrinos to enhance the hierarchy sensitivity of ATM data has
been mentioned in~\cite{c2m}.

\section{Concluding Remarks}

I have discussed the potential of future high statistics atmospheric
neutrino experiments, focusing on the possibility to resolve the
octant and the hierarchy ambiguities by exploring sub-leading
three-flavour effects.

The octant degeneracy in general is rather hard to resolve for
long-baseline experiments. In this case atmospheric data provide a
rather robust signature in the sub-GeV $e$-like events due to
oscillations with the solar frequency $\Delta m^2_{21}$. The
sensitivity is largely independent of the value of
$\theta_{13}$. Hence, if $\theta_{23}$ turns out to be non-maximal
atmospheric neutrinos provide a competitive method to resolve the
octant degeneracy.

\begin{figure}[!t]
\centering
  \includegraphics[width=0.45\textwidth]{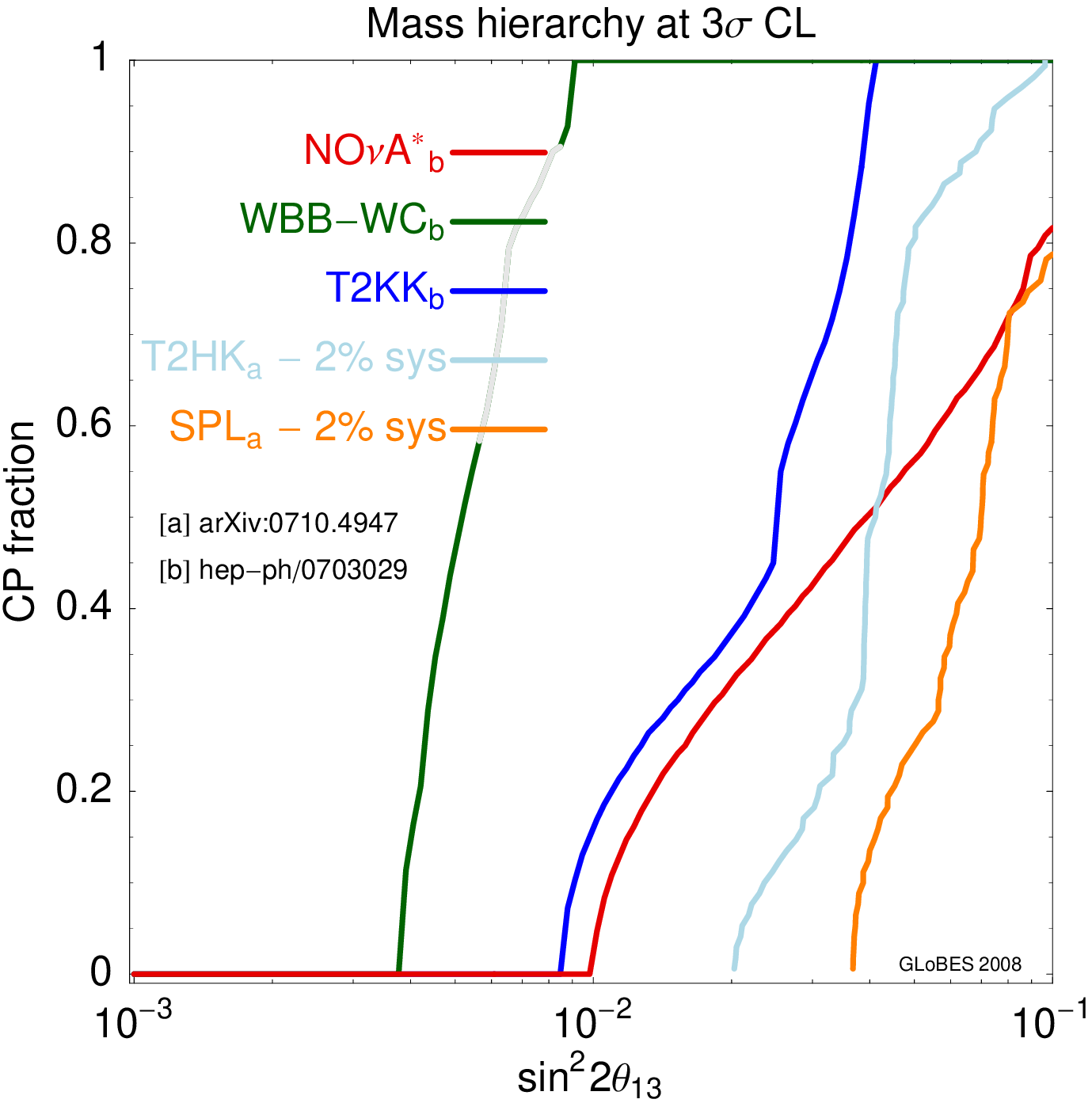}
\vspace*{-6mm}
  \caption{Fraction of $\delta_\mathrm{CP}$-values for which
  the neutrino mass hierarchy can be identified at $3\sigma$.
  SPL and T2HK include 5 Mt yr WC atmospheric neutrino data, while the
  other curves are for LBL data only. The setups
  are~\cite{Barger:2007jq} NO$\nu$A$^*$: 100~kt liquid Ar @ 820~km,
  3 yr $\nu$, 3 yr $\bar\nu$ @ 1.1~MW; T2KK: 270~kt WC @ 295 \& 1050
  km, 4 yr $\nu$, 4 yr $\bar\nu$ @ 4 MW; WBB: 300~kt WC @ 1290~km, 5yr
  $\nu$ @ 1 MW, 5yr $\bar\nu$ @ 2 MW.}
  \label{fig:sb}
\end{figure}

The determination of the neutrino mass hierarchy is based on the
resonant matter effect for neutrinos with long trajectories through
the Earth (large nadir angles) in the energy range of a few GeV. This
effect can be explored to determine the neutrino mass hierarchy if
$\theta_{13}$ is relative large. In this case data from atmospheric
neutrinos may be able to perform this measurment, or at least provide
complementary information to long-baseline data. However, if
$\sin^22\theta_{13}$ turns out to be less than a few$\times 0.01$ it
becomes exceedingly difficult for atmospheric neutrinos and the
sensitivity of long-baseline experiments with baselines $L\sim
1000$~km would be the preferred option to measure the neutrino mass
hierarchy. Fig.~\ref{fig:sb} shows that the ``short'' baseline
experiments SPL/T2HK + atmospheric data can determine the mass
hierarchy for $\sin^22\theta_{13} \gtrsim 0.05$ (at least for some
values of the CP phase), while, e.g., a wide band beam with $L \approx
1300$~km could provide solid sensitivity down to $\sin^22\theta_{13}
\simeq 0.008$.

Finally I mention the large potential of atmospheric neutrino data to
constrain (or eventually discover) all kinds of non-standard neutrino
properties. The reason for this good sensitivity is (in most cases)
the wide range in neutrino energies and baselines probed in
atmospheric neutrinos.  Some examples are non-standard interactions
\cite{nsi}, violation of Lorentz invariance~\cite{lorentz}, quantum
decoherence \cite{decoh}, neutrino decay \cite{decay}, or sterile
neutrinos \cite{sterile}.

\bigskip

{\bf Acknowledgments.} I thank the organizers for the very pleasant
atmosphere at the NOW2008 workshop and for financial support.  I
acknowledge the support of the European Community-Research
Infrastructure Activity under the FP6 ``Structuring the European
Research Area'' program (CARE, contract number RII3-CT-2003-506395).


\begin{thebibliography}{99}

\bibitem{SKatm}
  Super-K Collaboration,
  Y.~Fukuda {\it et al.}, 
  Phys.\ Rev.\ Lett.\  {\bf 81} (1998) 1562; 
%
  Y.~Ashie {\it et al.},
  Phys.\ Rev.\ D {\bf 71} (2005) 112005.


\bibitem{lbl}
  P.~Adamson {\it et al.}  [MINOS],
  Phys.\ Rev.\ Lett.\  {\bf 101} (2008) 131802;
%
  J.~Evans, these proceedings;
%
  D.S.~Ayres {\it et al.} [NOvA],
  hep-ex/0503053;
%
  M.~Goodman, these proceedings.

\bibitem{t2k}
  Y.~Itow {\it et al.} [T2K],
  hep-ex/0106019,
%
  H. Kakuno, these proceedings.

\bibitem{Huber:2004ug}
  P.~Huber {\it et al.}, 
  Phys.\ Rev.\ D {\bf 70} (2004) 073014
  [hep-ph/0403068].

\bibitem{Akhmedov:2008qt}
  E.~K.~Akhmedov, M.~Maltoni and A.~Y.~Smirnov,
  JHEP {\bf 0806} (2008) 072;
%
  JHEP {\bf 0705} (2007) 077;
%
E.~K.~Akhmedov, these proceedings.

\bibitem{laguna} D.~Autiero {\it et al.}, 
JCAP {\bf 11}, 011 (2007). 

\bibitem{Wcerenkov}
  C.~K.~Jung,
  hep-ex/0005046;
%
  K.~Nakamura,
  Int.\ J.\ Mod.\ Phys.\ A {\bf 18} (2003) 4053;
%
  A.~de Bellefon {\it et al.}, 
  hep-ex/0607026.

\bibitem{INO} G.~Rajasekaran,
  AIP Conf.\ Proc.\  {\bf 721} (2004) 243
  [hep-ph/0402246];
%
  S.~Goswami, these proceedings.

\bibitem{lar}
  A.~Ereditato and A.~Rubbia, 
  Nucl. Phys. Proc. Suppl. 154, 163 (2006). 

\bibitem{Choubey:2006jk}
  S.~Choubey,
  hep-ph/0609182.

\bibitem{Huber:2005ep}
  P.~Huber, M.~Maltoni, T.~Schwetz,
  Phys.\ Rev.\ D {\bf 71} (2005) 053006
  [hep-ph/0501037].

\bibitem{atm13}
  E.K.~Akhmedov {\it et al.}, 
  Nucl.\ Phys.\ B {\bf 542} (1999) 3;
%
  S.~T.~Petcov,
  Phys.\ Lett.\  B {\bf 434} (1998) 321;
%
  J.~Bernabeu, S.~Palomares Ruiz, S.T.~Petcov,
  Nucl.\ Phys.\ B {\bf 669} (2003) 255.

\bibitem{Gandhi:2007td}
  R.~Gandhi {\it et al.}, 
  Phys.\ Rev.\  D {\bf 76} (2007) 073012;
%
  R.~Gandhi {\it et al.}, 
  Phys.\ Rev.\  D {\bf 78} (2008) 073001.

\bibitem{Peres:2003wd}
  O.L.G.~Peres, A.Y.~Smirnov,
  Nucl.\ Phys.\ B {\bf 680} (2004) 479
  [hep-ph/0309312].

\bibitem{Gonzalez-Garcia:2004cu}
  M.C.~Gonzalez-Garcia, M.~Maltoni, A.Y. Smirnov,
  Phys.\ Rev.\ D {\bf 70} (2004) 093005.

\bibitem{c2m}
  J.E.~Campagne, M.~Maltoni, M.~Mezzetto and T.~Schwetz
  JHEP {\bf 0704}, 003 (2007) [hep-ph/0603172].

\bibitem{globes}
  P.~Huber, M.~Lindner, W.~Winter,
  Comput.\ Phys.\ Commun.\  {\bf 167} (2005) 195
  [hep-ph/0407333];
%
  P.~Huber {\it et al.}, 
  Comput.\ Phys.\ Commun.\  {\bf 177} (2007) 432
  [hep-ph/0701187].

\bibitem{Adamson:2005qc}
  P.~Adamson {\it et al.}  [MINOS Coll.],
  Phys.\ Rev.\ D {\bf 73}, 072002 (2006)
  [hep-ex/0512036].

\bibitem{Petcov:2005rv}
  S.~T.~Petcov and T.~Schwetz,
  Nucl.\ Phys.\ B {\bf 740}, 1 (2006)
  [hep-ph/0511277].

\bibitem{magn}
  T.~Tabarelli de Fatis,
  Eur.\ Phys.\ J.\ C {\bf 24}, 43 (2002);
%
  S.~Palomares-Ruiz and S.~T.~Petcov,
  Nucl.\ Phys.\ B {\bf 712} (2005) 392;
%
  D.~Indumathi and M.~V.~N.~Murthy,
  Phys.\ Rev.\ D {\bf 71} (2005) 013001;
%
  R.~Gandhi {\it et al.}, 
  Phys.\ Rev.\ D {\bf 73} (2006) 053001;
%
  S.~Choubey and P.~Roy,
  Phys.\ Rev.\ D {\bf 73}, 013006 (2006);
%
  A.~Donini {\it et al.},
  Eur.\ Phys.\ J.\  C {\bf 53} (2008) 599.




\bibitem{Huber:2008yx}
  P.~Huber and T.~Schwetz,
  Phys.\ Lett.\  B {\bf 669}, 294 (2008)
  [0805.2019].


\bibitem{Barger:2007jq}
  V.~Barger, P.~Huber, D.~Marfatia and W.~Winter,
  Phys.\ Rev.\  D {\bf 76}, 053005 (2007).

\bibitem{nsi}
  N.~Fornengo {\it et al.},
  Phys.\ Rev.\  D {\bf 65}, 013010 (2002);
%
  A.~Friedland, C.~Lunardini and M.~Maltoni,
  Phys.\ Rev.\  D {\bf 70}, 111301 (2004).
%


\bibitem{lorentz}
  G.~L.~Fogli {\it et al.},
  Phys.\ Rev.\  D {\bf 60} (1999) 053006;
%
  M.~C.~Gonzalez-Garcia and M.~Maltoni,
  Phys.\ Rev.\  D {\bf 70} (2004) 033010;
%
  M.~C.~Gonzalez-Garcia, F.~Halzen and M.~Maltoni,
  Phys.\ Rev.\  D {\bf 71}, 093010 (2005).


\bibitem{decoh}
  G.~L.~Fogli {\it et al.}, 
  Phys.\ Rev.\  D {\bf 67}, 093006 (2003).

\bibitem{decay}
  G.~L.~Fogli {\it et al.}, 
  Phys.\ Rev.\  D {\bf 59}, 117303 (1999);
%
  S.~Choubey and S.~Goswami,
  Astropart.\ Phys.\  {\bf 14}, 67 (2000);
%
  V.~D.~Barger et al., 
  Phys.\ Lett.\  B {\bf 462}, 109 (1999);
%
  M.~C.~Gonzalez-Garcia and M.~Maltoni,
  Phys.\ Lett.\  B {\bf 663}, 405 (2008).


\bibitem{sterile}
  S.~Choubey,
  JHEP {\bf 0712}, 014 (2007).


\end{thebibliography}
\end{document}